\documentclass[twocolumn,aps,prd,showpacs,superscriptaddress,floats,floatfix,showkeys,notitlepage]{revtex4-1}

\usepackage{orcidlink}
\usepackage{lipsum}
\usepackage{graphicx}
\usepackage{subfigure}
\usepackage{palatino}
\usepackage{changes}
\usepackage{hyperref}
\hypersetup{colorlinks=true,linkcolor=blue,urlcolor=blue,citecolor=blue}
\usepackage[toc,page]{appendix}
\usepackage[normalem]{ulem}
\usepackage{adjustbox}
\usepackage{latexsym}
\usepackage{amsmath}
\usepackage{amssymb}
\usepackage{amsfonts}
\usepackage{times}
\usepackage{dcolumn}
\usepackage{bm}
\usepackage{tikz}
\usepackage{bigints}
\usepackage{array,tabularx,multirow}
\usepackage[tracking=true]{microtype}
\SetTracking{}{500}
\SetTracking{encoding={*}, shape=sc}{40}
\UseRawInputEncoding 
\allowdisplaybreaks

\usepackage{braket}
\usepackage{tensor}
\usepackage{slashed}
\usepackage{booktabs} 
\usepackage{float}
\newcommand\aap{Astron. Astrophys.}                
\newcommand\mnras{Mon. Not. Roy. Astron. Soc.}             
\newcommand\apjl{Astrophys. J. Lett.}                

\begin{document}

\title{Distinguishability of a naked singularity from a black hole in dynamics and radiative signatures}

\author{Indu K. Dihingia}
\email{ikd4638@sjtu.edu.cn, ikd4638@gmail.com}
\affiliation{Tsung-Dao Lee Institute, Shanghai Jiao-Tong University, Shanghai, 520 Shengrong Road, 201210, People's Republic of China}

\author{Akhil Uniyal}
\email{akhil\_uniyal@sjtu.edu.cn}
\affiliation{Tsung-Dao Lee Institute, Shanghai Jiao-Tong University, Shanghai, 520 Shengrong Road, 201210, People's Republic of China}

\author{Yosuke Mizuno}
\email{mizuno@sjtu.edu.cn}
\affiliation{Tsung-Dao Lee Institute, Shanghai Jiao-Tong University, Shanghai, 520 Shengrong Road, 201210, People's Republic of China}
\affiliation{School of Physics \& Astronomy, Shanghai Jiao-Tong University, Shanghai, 800 Dongchuan Road, 200240, People's Republic of China}
\affiliation{Institut f\"{u}r Theoretische Physik, Goethe Universit\"{a}t, Max-von-Laue-Str. 1, 60438 Frankfurt am Main, Germany}
\date{\today}

\begin{abstract}
Can a naked singularity (NkS) be distinguished from a black hole (BH)? We have investigated it with cutting-edge general relativistic magneto-hydrodynamic (GRMHD) simulations, followed by general relativistic radiation transfer (GRRT) calculation for magnetized accretion flow around NkS and BH. Based on our simulations, the accreting matter close enough to the singularity repels due to effective potential. This prevents matter from reaching a NkS and forms a quasi-spherical symmetric density distribution around it, unlike the accretion flows around a BH. We observe an order of magnitude higher mass flux through the jet and much stronger wind from a NkS than a BH. We found that the jet launching mechanism in a NkS differs significantly from that in a BH. In the horizon-scale images, a NKs shows a photon arc instead of a photon ring that is shown around a BH. In summary, the flow dynamics and radiative properties around an NkS are distinctly different from a BH. These properties would be useful to either confirm or rule out such exotic compact objects through future observations.
\end{abstract}

\maketitle

\date{\today}


\section{Introduction} \label{sec:intro}
A Naked Singularity (NkS) is a fascinating theoretical object that astronomers are constantly trying to investigate possibilities of its existence. If cosmic censorship is assumed, the formation of a NkS is forbidden \citep{Penrose1969,Penrose2002}. There are different kinds of a NkS proposed in the literature; most of them are from the beyond general relativity (GR) \cite[e.g.,][]{Kovacs-Harko2010,Vieira-etal2014,Goluchova-etal2015,Boshkayev-etal2016}. However, NkS solutions also appear in GR, such as the Chazy-Curzon solution with $Q>M_*$, i.e., the Reissner-Nordstr{\"o}m (RN) solution, and the ``superspinars'' $a_*>M_*$ Kerr naked singularity solution \citep{Griffiths2009}, where $Q$ is electric charge, $M_*$ is a mass, and $a_*$ is normalized spin .

This study investigates a NkS within GR, i.e., ``superspinars'', although they might not be stable under small perturbations. Previous studies have shown that certain objects with rapid rotation in dimensions of $3+1$ and beyond can be unstable \citep{Emparan-etal2003,Cardoso-etal2008}. Recent analytical studies suggest that the stability of such objects depends on the boundary conditions at the surface, and certain classes of such objects are stable \citep{Zhong-eatl2023}. At this point, it is impossible for us to study these aspects of a NkS, and here we consider that it to be stable. Previously, limited studies have been done on accretion flows around a NkS under the hydrodynamic assumption. It has been shown that in a NkS, the accretion flows cannot reach the central NkS and accumulate around it, forming a high-density cloud \cite{Bambi2009, Dihingia-etal2020a}. Infect, the study also suggests that accretion flows around a NkS can cloak itself as an opaque (quasi) spherical surface of matter for a distant observer \citep{Vieira-etal2023}. 

Recent hydrodynamic simulations by \cite{Kluzniak-Krajewski2024} show that the accretion flows around a NkS (in RN spacetime) form a toroidal inner structure and strong outflows. Thus, jets and outflows from an NkS could provide insights into a NkS, as well as distinguishable features from black holes (BH). Accordingly, in this study, we consider plasma dynamics around both a NkS and a BH and perform general relativistic magneto-hydrodynamic (GRMHD) simulations. From an extensive analysis of flow properties and comparison, we discuss their distinguishability. 

The last few decades have been dedicated to calculating the horizon-scale images of BHs, which were finally observed by the Event Horizon telescope (EHTC) \citep{EHTC2019,EHTVII-etal2021,EHTC2022,EHT2024,Raymond-etal2024}. The image of a BH depicts the shadow surrounded by an asymmetric ring of emission. This emission is due to the gravitational lensing of light from hot, accreting matter, highlighting the strong-field effects of GR and the Doppler beaming caused by the rotation of the accretion flows. Since a NkS does not have an event horizon, the prograde orbits spiral into the central singularity. On the other hand, the retrograde orbits cast an arc-shaped shadow with a central dark spot \cite{Kenta-etal2009, Kumar-etal2021,Nguyen-etal2023}. To our knowledge, no one has yet calculated the image of a NkS using a realistic fluid background and radiation processes. This work will be the first to calculate the horizon-scale images through the application of general relativistic radiation transfer (GRRT) on realistic GRMHD simulations of magnetized accretion flows onto a NkS.

\section{Numerical setup}\label{sec:math}
This study examines accretion flow using ideal axisymmetric and 3D GRMHD simulations by the \texttt{BHAC} code \citep{Porth-etal2017, Olivares-etal2019} in Modified Kerr-Schild coordinates around a Kerr BH and a NkS. Usually, ``inflow'' boundary conditions are used at the inner boundary (inside the outer horizon but outside the inner horizon) to simulate accretion flows around BHs, where the outflow of matter is prohibited at the inner boundary. Around the NkS, things are unclear. We set two types of inner boundary conditions: (1) identical to BH (``inflow'') and (2) ``reflective''. We use the ``reflective'' boundary condition to make the flow bounce back with opposite radial velocity. A detailed description of the numerical setup is given in Appendix~\ref{appendixA}.

\section{Dynamical difference between a NkS and a BH}\label{sec4}
\begin{figure*}
    \centering
    \includegraphics[width=0.85\textwidth]{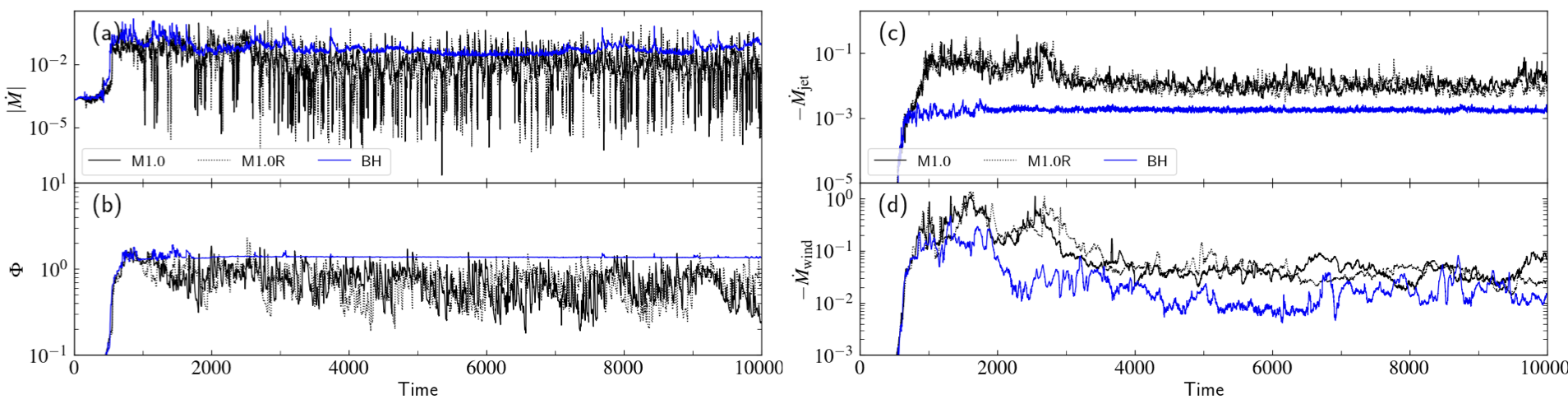}
    \caption{Left panels ({\it a,b}) show the profiles for accretion rate ($|\dot{M}|$, magnitude only) and magnetic flux ($\Phi$); right panels ({\it c,d}) show the mass outflow rate through jet ($-\dot{M}_{\rm jet}$) and wind ($-\dot{M}_{\rm wind}$) for both BH and NkS cases.}
    \label{fig:Fig-02}
\end{figure*}

In this section, we consider two cases from NkS simulations (M$1.0$ and M$1.0$R) with the same inner boundary location, i.e., $r_{\rm in, edge}=\exp(-1)\,r_g \simeq 0.368\,r_g$ and compare with the simulation around a BH. Note that simulation results converge if the inner boundary is close enough to the singularity and the impacts of inner boundary conditions vanish (see Appendix~\ref{AppendixB} for details). First, we calculate the mass accretion rate ($\dot{M}$) and magnetic flux ($\Phi$) at $r=1.141\,r_g$ (horizon of the BH case) for all the cases, where they are defined as $\dot{M}=2\pi \int \sqrt{-g}\rho u^r d\theta$ and $\Phi= \pi\int \sqrt{-g} |B^r| d\theta $, respectively. The time evolution of these two quantities is shown in Fig.~\ref{fig:Fig-02}(a) and (b). The accretion rates for a NkS have positive as well as negative values due to the onset of outflow/jet. Accordingly, in the panels, we show only the magnitude of the accretion rate ($\dot{M}$). Along with these two quantities, we also show the mass outflow rate through the jet ($\dot{M}_{\rm jet}$) and wind ($\dot{M}_{\rm wind}$) at $r=50\,r_g$ in Fig.~\ref{fig:Fig-02}(c) and (d) following the same definitions as \cite{Nathanail-etal2020}.

Both the accretion rate and the magnetic flux reach a quasi-steady state for NkS as well as BH cases after $t=2000\,t_g$. Due to the reflective nature of the NkS, the average value of the accretion rate is much smaller than that of the BH. For instance, within time $t=8000-10000\,t_g$, the average value of the accretion rate of NkS and BH cases is $\langle\dot{M}\rangle=0.0178$ and $\langle\dot{M}\rangle=0.0670$, respectively. Note that if the radius of calculation is taken further inside for NkS, the accretion rate decreases further. Similar to the accretion rate profiles, we also observe distinct differences between magnetic flux profiles in the case of the NkS and the BH in Fig.~\ref{fig:Fig-02}(b). The BH's magnetic flux is much higher than that of the NkS. The average value of the magnetic flux of NkS and BH cases is $\langle\Phi\rangle=0.62$ and $\langle\Phi\rangle=1.36$, respectively. This suggests that NkS is inefficient for the accumulation of magnetic fields around it due to its reflective nature (see Appendix~\ref{appendixC}). The impact of the reflective nature of the NkS is also visible in the jet and wind outflow rates seen in Fig.~\ref{fig:Fig-02}(c) and (d). We observe an order of magnitude higher jet outflow rate for the NkS than that of the BH. The values of jet outflow rates during simulation time $t=8000-10000,t_g$ are as follows: $\dot{M}_{\rm jet}=-0.0134$ (M$1.0$), $\dot{M}_{\rm jet}=-0.0093$ (M$1.0$R), and $\dot{M}_{\rm jet}=-0.0017$ (BH). From the wind outflow rates ($\dot{M}_{\rm wind}$), a NkS has a higher value than those of the BH. The difference between a NkS and a BH approximately doubles, which is not drastically higher. This is because the wind is primarily driven by the accretion flows, not the central object itself. 

\begin{figure*}
    \centering
    \includegraphics[width=0.85\textwidth]{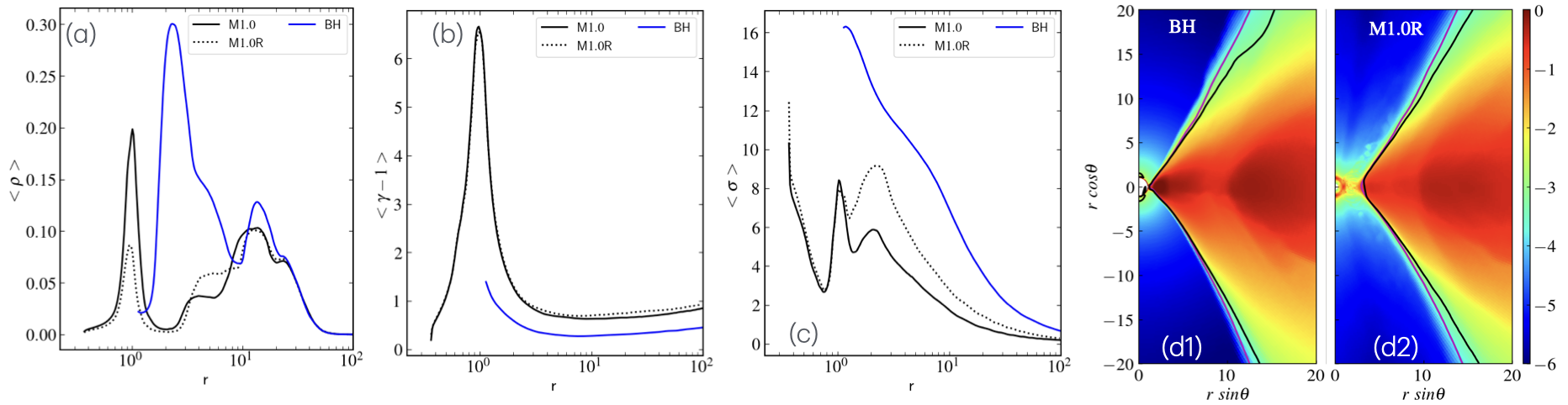}
    \caption{Time-averaged vertically averaged density $\langle \rho \rangle$, Lorentz factor $\langle \gamma -1 \rangle$, and magnetization $\langle\sigma\rangle$ for different simulation models. (d1) and (d2) show the time-averaged density distribution for models M$1.5$ and M$1.5$R. See text for more detail.
    }
    \label{fig:Fig-03}
\end{figure*}

Next, we compare flow properties between NkS and BH. To do that, we plot time-averaged ($t=8000-10000\,t_g$) and vertically averaged density ($\langle \rho \rangle$), Lorentz factor ($\langle \gamma -1 \rangle$), and magnetization ($\langle \sigma \rangle$) for different simulation models in panels Fig.~\ref{fig:Fig-03}(a), (b), and (c), respectively. We also show the time-averaged density distribution for models BH and NkS M$1.0$R in panels Fig.~\ref{fig:Fig-03}(d1) and (d2), respectively. Magenta and black contours in these two panels correspond to $\sigma=1$ and $-hu_t=1$, respectively, where $-hu_t$ is Bernoulli parameter, $h$ is specific enthalpy, and $u_t$ is time component of covariant four-velocity. In Fig.~\ref{fig:Fig-03}(a), we see that the density close to the event horizon is small for BH. However, for a NkS, we have an additional peak around $r\sim r_g$. This means that there is a so-called zero gravity surface around a NkS (see Appendix~\ref{appendixC}). On the zero gravity surface, matter accumulates and creates a spherically symmetric mass distribution around a NkS. This difference can also be seen in the poloidal density distribution in panels Fig.~\ref{fig:Fig-03}(d1) and (d2). In panel Fig.~\ref{fig:Fig-03}(b), the Lorentz factor is the highest at the horizon for a BH. However, for a NkS, as the flow approaches a singularity, the Lorentz factor increases and reaches a peak value around $r\sim r_g$. Furthermore, as the flow reaches very close to singularity, the Lorentz factor drops. Previous semi-analytical studies also reported such behaviors of the accretion flow around a NkS \citep{Dihingia-etal2020a, Vieira-etal2023}. Subsequently, far from the central object, the Lorentz factor is much smaller for a BH as compared to a NkS due to the opposite nature of effective potential (attractive/repulsive). For the same reason, the magnetic flux can accumulate efficiently in a BH but not in a NkS. As a result, the magnetization is higher around a BH than that of a NkS (Fig.~\ref{fig:Fig-03}(c)). Furthermore, in Fig.~\ref{fig:Fig-02}(d1) and (d2), we see that the $\sigma=1$ contour is rooted to the horizon for the BH case, but for the NkS case, it is not rooted to the inner boundary or the zero gravity surface. This suggests that the matter around the NkS is magnetically dominated despite the lower magnetic flux accumulated around it.

\begin{figure*}
    \centering
    \includegraphics[width=0.80\textwidth]{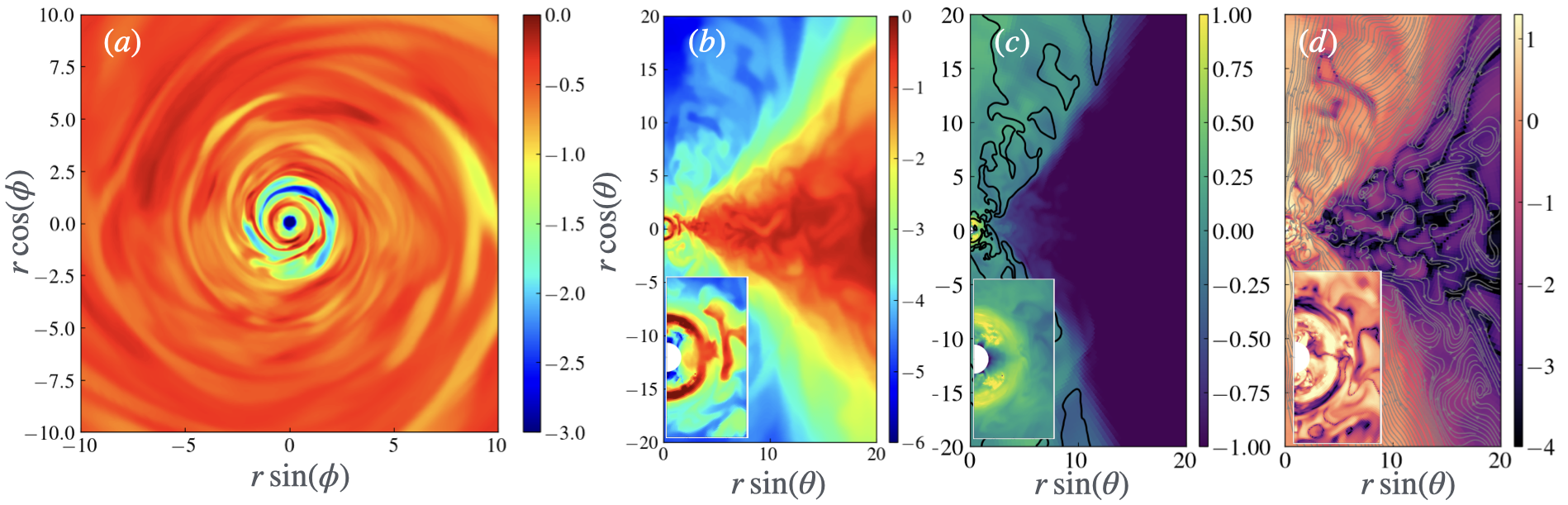}
    \caption{(a) Density distribution on the equatorial plan (b) on the $\phi=0$ plan, (c) Lorentz factor ($\gamma -1$), and (d) magnetization at simulation time $t=2000\,t_g$ for model M$1.0$R3DHR. The inset of panels (b,c, and d), the same is shown close to the NkS. The black lines in panel (c) and the gray lines in panel (d) correspond to $\sigma=1$ and the poloidal magnetic field lines.}
    \label{fig:Fig-06}
\end{figure*}

Moreover, we test our findings of dynamical flow properties in a NkS with full 3D simulations. Due to numerical limitation, we only show results for model M$1.0$R3DHR at time $t=2000\,t_g$.
Accordingly, Fig.~\ref{fig:Fig-06} shows logarithmic (a) density distribution on the equatorial plane, (b) on the $\phi=0$ plane, (c) Lorentz factor ($\gamma -1$), and (d) magnetization with a zoomed version in the inset of panels (b-d). We find exactly similar results to those from the axisymmetric simulations. These panels also clearly demonstrate the formation of density distribution around a NkS. However, the density on the equatorial plane suggests the formation of density spirals, which are not seen in axisymmetric cases. The distribution of the Lorentz factor shows the negligible value at the boundary and the highest value around $r\sim r_g$, clearly showing the relativistic jet around the rotation axis of the NkS. The jet region has higher magnetization. Interestingly, the magnetization close to the NkS has a lower value as in the axisymmetric model with a smaller inner boundary. We note that the same results for a low-resolution case at a much later simulation time ($t=8000\,t_g$) are shown in Fig.~\ref{fig:Fig-A3} and discussed in Appendix~\ref{appendixD}.

\section{Jet launching from a NkS}\label{sec6}
In previous sections, we have observed a relativistic jet from a NkS. However, the jet launching mechanism is not clear. In the case of a rotating BH, one of the promising jet launching mechanisms is well-known, i.e., the Blandford-Znajek (BZ) mechanism \cite{Blandford-Znajek1977}. The BZ mechanism requires an ergosphere and poloidal magnetic field lines rooted in it. For a NkS, we do not see magnetic field lines rooted close to the central object (Fig.~\ref{fig:Fig-06}d). In fact, the magnetic field configuration near the singularity is quite turbulent due to the reflection of the matter. However, jet Lorentz factor is $\gamma\simeq4-6$ ($\log(\gamma -1)\simeq 0.5-0.75$) at the radius $r\sim 10\,r_g$ (Fig.~\ref{fig:Fig-06}c). The Lorentz factor close to the singularity is very high $\gamma\sim 10$ (Fig.~\ref{fig:Fig-06}c). It suggests that the reflection of falling matter by the potential barrier launches the jet. Around the jet base, the Lorentz factor decreases, indicating the dissipation of energy during propagation through the medium. Furthermore, it is accelerated by magnetic fields, as shown in panel Fig.~\ref{fig:Fig-03}(b), where we observe a slight increase in the Lorentz factor as matter moves from $r=10\,r_g$ to $r=100\,r_g$. On the contrary, the jet launched via the BZ mechanism starts with a very small Lorentz factor, and the value of the Lorentz factor increases as the jet moves far from the BH \cite[e.g.,][]{Dihingia-etal2021}. Therefore, the jet formation mechanism between a BH and a NkS is clearly different.

\section{Radiative distinguishability between NS and BH}\label{sec7}
\begin{figure}
    \centering
    \includegraphics[width=0.49\textwidth]{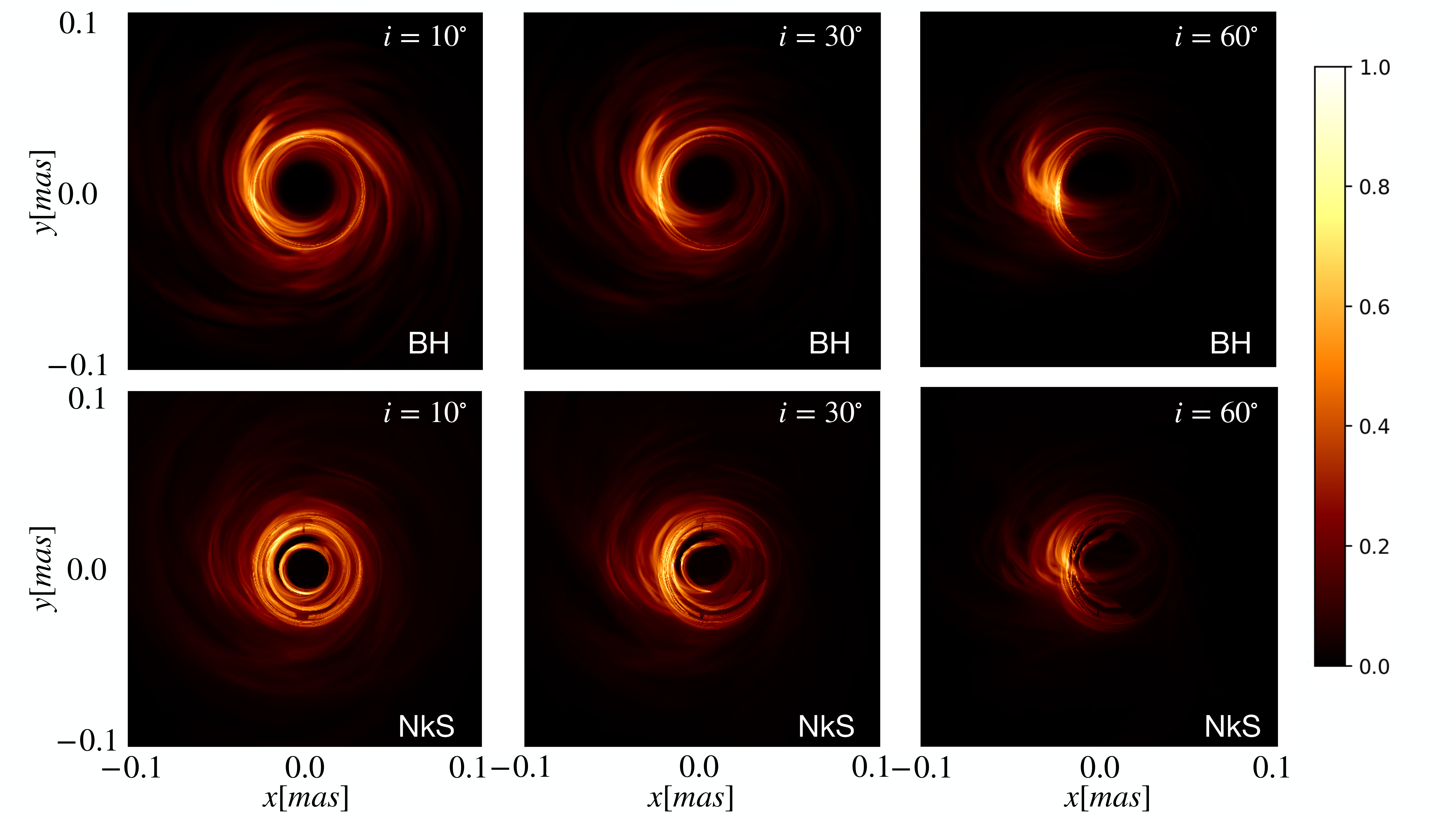}
    \caption{Intensity map for a BH (upper) and a NkS (lower) at three different inclination angle at 230 GHz}.
    \label{fig:Fig-05}
\end{figure}

In this section, we used GRRT code {\tt RAPTOR} \cite{Bronzwaer:2018lde, Bronzwaer:2020kle} to calculate the horizon scale images of the two compact objects. We fixed the observed frequency at 230 GHz and changed the inclination angle by 10, 30, and 60 degrees for clear visibility of the inner structure of the image at simulation time $t=2000\,t_g$ to observe the differences as shown in Fig.~\ref{fig:Fig-05}. For the calculation of the images of black holes with Kerr parameter $a_*=0.9375$. For NkS, we calculate it from the model M$1.0$R3DHR with spin parameter $a_*=1.01$. The grid resolution in the image plane is $600*600$ pixels covering the field of view (FoV) $0.25\, mas^2$ $(40 \, r_g^2)$. Note that in the Fig.~\ref{fig:Fig-05}, we show the images within $\pm 0.1\,mas$ FoV for clear visibility. The target source is considered Sgr~A$^*$ having the mass $M=4.14 \times 10^6$ \(\textup{M}_\odot\) and distance $8.127$ kpc \citep{EventHorizonTelescope:2022wkp}. For comparison, we only considered thermal synchrotron radiation in both cases. Here, for simplicity, we consider a single temperature approximation to calculate the electron temperature, i.e., $T_e=T_i$. For a NkS, we fixed the geodesic integration cutoff at the position of the inner boundary used in the GRMHD simulations, i.e., $r_{\rm cut}\simeq 0.368\, r_g$. Ideally, the inner boundary can be taken even closer for solving the geodesics. However, since the matter close to the boundary is negligible (so the radiation) and therefore taking it inside the $r_{\rm in, edge}$ will not alter the image of the NkS.
For the BH, it is the outer horizon position. For this reason, we see a dark region in the middle of the images; however, for the NkS, we also see some arc-like structures before reaching the inner boundary position. This feature is uniquely observed for the NkS in the previous studies by the analytical calculations \cite{Felice1975, Kumar-etal2021, Tavlayan:2023vbv}. In the case of the BH, we see the complete photon ring, which clearly distinguishes the NkS from the BH. Additionally, the NkS images formed some emission rings before reaching the inner boundary (clearly visible at a lower inclination angle) because of the matter accumulation as shown in Fig. \ref{fig:Fig-06}. These kinds of features can only be seen due to the existence of such a zero-gravity surface near the naked singularity. 

\section{Conclusions and discussion}\label{concl}
In this study, we performed axisymmetric and 3D GRMHD simulations around a NkS and BH. Due to reflective environment near NkS, very near the NkS remain inaccessible for accretion flows, which was reported by earlier semi-analytic or hydrodynamic simulations \citep{Bambi2009,Dihingia-etal2020a,Vieira-etal2023,Kluzniak-Krajewski2024}. As a result, we observe that the mass outflow through the jet and wind is higher for a NkS than it is for a BH. In fact, the outflow through the jet is significantly higher for a NkS compared to a BH. Unlike BH, A NkS forms a quasi-spherically symmetric density surface very close to the singularity. Similar features were also suggested by earlier semi-analytic/hydrodynamic studies \citep{Bambi2009,Vieira-etal2023}. The Lorentz factor is highest at the event horizon for a BH; however, the Lorentz factor decreases to a very small value as accretion flows approach to a NkS. The jet launching mechanism for the NkS differs from that for the BH. We did not find poloidal magnetic field lines rooted in NkS. Near NkS, the magnetic field lines are very turbulent. The flow properties suggest that the jet is launched due to the bounce of the falling of the relativistic matter on the potential barrier.
    
Far from the central objects, the shadow images are quite similar for both BH and NkS. However, the shadows they cast closer are clearly different in the GRRT images. The horizon-scale images of a NkS cast an arc-like structure instead of a ring that is seen in a BH, as noted by earlier analytical calculations \citep[e.g.,][]{Felice1975,Kenta-etal2009,Kumar-etal2021, Tavlayan:2023vbv,Nguyen-etal2023}.

In principle, we could use these features in the future to study the evidence for a NkS. A recent study on the images from M\,87 and Sgr\,A$^*$ ruled out most of the known NkS solutions \citep{Broderick-Salehi2024}. This does not imply that a NkS is either impossible or feasible to exist. The existence of the extra density distribution near the NkS may have other observational implications; such as, it can have some impacts on the variability of the light curves or the spectral features of the source.

This study is performed with a very simple ``superspinner'' ($a_*>1$) type singularity that has a smaller photon sphere than that of a BH and zero-gravity surface. Although many other spacetimes can have a NkS for given ranges of parameters \citep[e.g.,][]{Reissner1916,Janis-etal1968,Yajima-etal2001,Joshi-etal2011,Astorino2013,Joshi-etal2014,Banik-etal2017,Kocherlakota-Rezzolla2020,Fernandes2020,Chen-etal2024,Viththani-etal2024}, depending on the complexities of the spacetime, the flow dynamics and radiative properties may be different. An extensive comparison study in the future is needed.

\begin{acknowledgements}
This research is supported by the National Key R\&D Program of China (No.\,2023YFE0101200), the National Natural Science Foundation of China (Grant No.\,12273022), and the Shanghai Municipality orientation program of Basic Research for International Scientists (Grant No.\,22JC1410600). The simulations were performed on the TDLI-Astro cluster in Tsung-Dao Lee Institute, Pi2.0, and Siyuan Mark-I clusters in the High-Performance Computing Center at Shanghai Jiao Tong University.
This work has made use of NASA's Astrophysics Data System (ADS).
\end{acknowledgements}

\appendix
\section{Model parameters}\label{appendixA}
We use spherical polar coordinates $(r, \theta, \phi)$ with logarithmic spacing in the radial direction ($r$, up to $r=2500\,r_g$) and uniform spacing in the polar and azimuthal directions. To investigate the effect of inner boundary conditions, we choose different locations for inner boundaries ($r_{\rm in, edge}$), they are given in table~\ref{tab-01}. We run the simulations in a standard framework of units with $G=M_*=c=1$. Here, $G$ and $c$ represent the universal gravitational constant and the speed of light, respectively. In this unit system, length and times are given in units of $r_g=GM_*/c^2$ and $t_g=GM_*/c^3$. In this study, we perform axisymmetric simulations with $512\times256$ grid resolution for comparisons between flow properties.  Moreover, we perform two 3D simulations, one in lower resolution ($256\times80\times64$) and the other in higher resolution ($512\times160\times128$) to test the consistency of our results from axisymmetric runs. This study mainly focuses on the differences in the accretion flows around a BH and a NkS. Therefore, we choose the Kerr metric with two spin parameters: $a_*=0.99$ for a BH and $a_*=1.01$ for a NkS.

We set the initial conditions for the simulation using the FM torus \citep{Fishbone-Moncrief1976,Uniyal-etal2024} with the inner edge of the torus $r_{\rm in}=10$. We choose the radius of density maximum for the BH case to be $r_{\rm max}=18.0$ and for the NkS case $r_{\rm max}=18.6$. This difference in $r_{\rm max}$ is needed to make the mass contained within the initial torus the same for both cases. Additionally, a non-zero azimuthal component of the vector potential $A_\phi \propto \max(\rho/\rho_{\rm max}-0.2,0)$ is supplied initially. To ensure divergence-free magnetic field throughout the simulation domain, the constrained-transport method is utilised during simulation \citep[see][]{Olivares-etal2019}.
Finally, the strength of the magnetic field is provided by supplying the minimum value of the initial plasma-$\beta$, which we choose the same for all the simulation models, i.e., $\beta_{\rm min}=100$. All other details of the simulation models are displayed in table~\ref{tab-01}. Note that we do not run our simulations for a very long time. Because our goal is to understand only the flow structure very close to the central object ($r\lesssim 5\,r_g$ or so). This region comes to a quasi-steady state within $t\sim 1500\,t_g$.

\begin{table}
\centering
\caption{Table displays the model name, locations of inner boundary in units of $r_g$($r_{\rm in, edge}$), effective resolution (SMR level), maximum simulation time in units of $t_g$, and the inner boundary condition, respectively.  }
  \begin{tabular}{| c | c | c | c | c |}
    \hline
     Model & $r_{\rm in,edge}$& Eff. Res.$^n$ (L) & Time & Bound.\\ 
    \hline
    BH &   $1.0$  & $512\times256$ (1) & 5000 & ``inflow''\\
     M$0.5$   & $0.6065$ & $512\times256$ (1) & 5000 & ``inflow'' \\
     M$0.5$R  & $0.6065$ & $512\times256$ (1) & 5000 & ``reflective'' \\
     M$1.0$   & $0.3680$ & $512\times256$ (1) & 10000 & ``inflow'' \\
     M$1.0$R3DLR   & $0.3680$ & $256\times80\times64$ (1) & 5000 & ``reflective'' \\
     M$1.0$R3DHR   & $0.3680$ & $512\times160\times128$ & 2000 & ``reflective'' \\
     M$1.0$R   & $0.3680$ & $512\times256$ (1) & 10000 & ``reflective'' \\
     M$1.5$   & $0.2231$ & $512\times256$ (1) & 5000 & ``inflow'' \\
     M$1.5$R   & $0.2231$ & $512\times256$ (1) & 5000 & ``reflective'' \\
    \hline
  \end{tabular}
\label{tab-01}
\end{table}

\begin{figure*}
    \centering
    \includegraphics[width=0.9\textwidth]{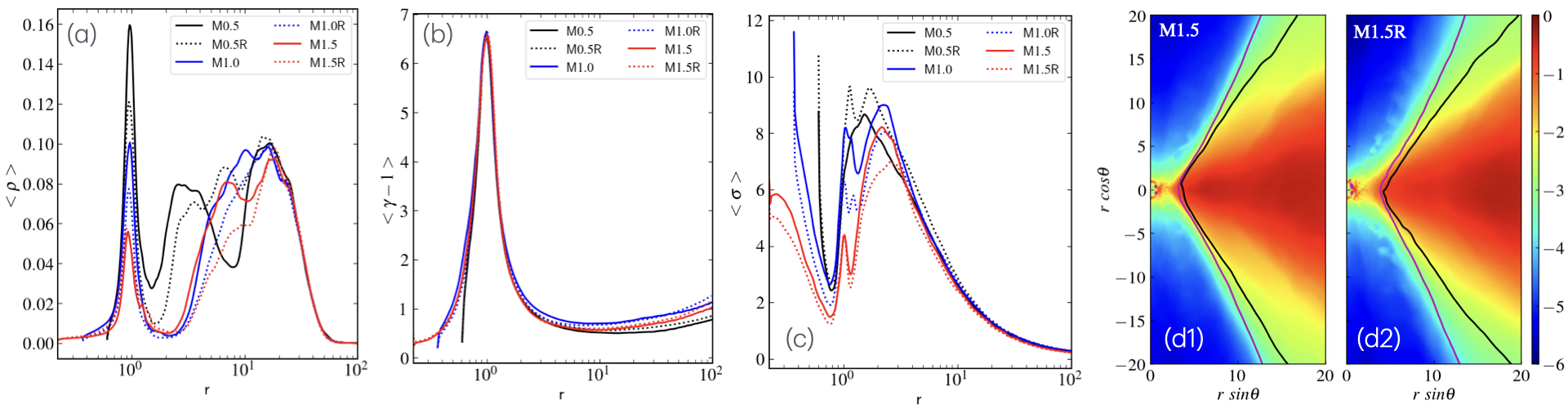}
    \caption{Same as Fig.~\ref{fig:Fig-03}, but comparison is shown for a different inner edge. Time averaging is performed at $t=4000-5000\,t_g$. }
    \label{fig:Fig-01}
\end{figure*}

\section{Inner boundary effects}\label{AppendixB}
Here, we study the impacts of the inner boundary conditions for NkS cases on the simulation results. To do that, we consider three inner boundary locations with two inner boundary conditions (``inflow'' (M$0.5$, M$1.0$, M$1.5$) and ``reflective'' (M$0.5$R, M$1.0$R, M$1.5$R)). To compare the results, we plot time-averaged ($t=4000-5000\,t_g$) vertically averaged density ($\langle \rho \rangle$), Lorentz factor ($\langle \gamma -1 \rangle$), and magnetization ($\langle \sigma \rangle$) for different simulation models in panels of Fig.~\ref{fig:Fig-01}(a), (b), and (c), respectively. We also show the time-averaged density distribution for models M$1.5$ and M$1.5$R in panels of  Fig.~\ref{fig:Fig-01}(d1) and (d2), respectively. Magenta and black contours in these two panels correspond to $\sigma=1$ and $-hu_t=1$, respectively. In the panels (Fig.~\ref{fig:Fig-01}(a,b,c)), solid and dotted lines correspond to ``inflow'' and ``reflective'' boundary conditions. 

For all the models, we see a strong density peak around radius $r\sim r_g$ apart from the matter in the body of the accretion flow ($2\le r/r_g\le 50$). The difference in the magnitude of the peak for ``inflow'' and ``reflective'' boundary conditions becomes small with decreasing the inner boundary radius (Fig.~\ref{fig:Fig-01}a). These results suggest that as long as the inner boundary is close enough to the singularity, the effect of the boundary condition reduces. On the contrary, the peak values for the Lorentz factor are quite similar for all the cases (Fig.~\ref{fig:Fig-01}b). The differences in the density peaks for $r_{\rm in, edge}=0.6065, 0.3680,~{\rm and}~ 0.2231\,r_g$ are given by $|\Delta \rho_{\rm max}|\sim 0.042,0.025,~{\rm and}~0.005$, respectively. This suggests convergence of results with decrease in inner boundary of the simulation domain.
It peaks around the density maximum $r\sim r_g$, and as the flow approaches the NkS, the Lorentz factor decreases. For the model with the innermost inner boundary location (M$1.5$ and M$1.5$R), we observe a smooth decrease in the Lorentz factor. However, for other models, very sharp drops are seen at the inner edge. Finally, the value of magnetization ($\langle \sigma \rangle$) shows an increasing trend as the flow approaches from the density peak to the NkS for larger inner boundary locations (M$0.5$, M$0.5$R, M$1.0$, M$1.0$R). However, for innermost inner boundary cases, we observe a saturation value at the boundary (Fig.~\ref{fig:Fig-01}c). This behavior could provide a reasonable description of the magnetic properties of a NkS. Eventually, the time-averaged density distribution on the poloidal plane for models (M$1.5$ and M$1.5$R) in panels of Fig.~\ref{fig:Fig-01}(d1) and (d2) confirms it. We cannot find any distinguishable differences between them. The contours $\sigma=1$ and $-hu_t=1$ that define jet and wind can also be seen in quite similar orientations. In summary, the inner boundary conditions (``inflow'' or ``reflective'') do not matter if the inner boundary is close enough to the singularity. 

\begin{figure}[h]
    \centering
    \includegraphics[width=0.45\textwidth]{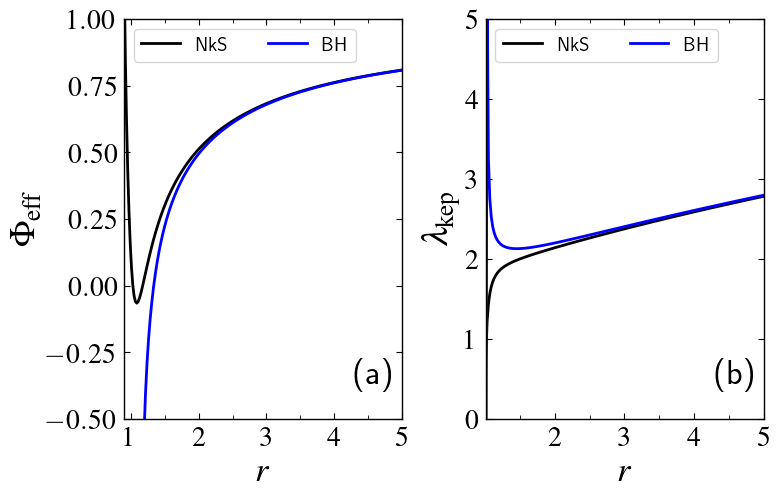}
    \caption{Plot of effective potential (a) and Keplerian angular momentum (b) on the equatorial plane for NkS ($a_*=1.01$) and BH ($a_*=0.99$).}
    \label{fig:Fig-A1}
\end{figure}

\section{Effective potential and Keplerian angular momentum around NkS and BH}\label{appendixC}
To understand the basic differences between a NkS and a BH, in Fig.~\ref{fig:Fig-A1}(a), we show the effective potential ($\Phi_{\rm eff}$) experienced by the fluid on the equatorial plane ($\theta=\pi/2$) for specific angular momentum $\lambda=-u_\phi/u_t=1.65$ for both central objects. The expression for the effective potential experienced by a fluid element is taken from \cite{Dihingia-etal2018a}. Note that, for a BH, the spin parameter is $a_*=0.99$, whereas for a NkS it is given by $a_*=1.01$. We observe that the potential has a positive gradient for a BH, making it attractive in nature. It is always attractive near the horizon, irrespective of the specific angular momentum of the fluid. On the other hand, the effective potential near the NkS is repulsive, although it was attractive up to $r=1.88\,r_g$ for this particular case, which acts as a zero gravity radius. Note that the zero gravity radius depends on the angular momentum of the flow. Fig.~\ref{fig:Fig-A1}(b) shows the Keplerian angular momentum on the equatorial plane ($\lambda_{\rm kep}=(a_*^2-2 a_* x+x^4)/(a_*+x^3-2 x); ~~ x=r^{1/2}$, \cite{Novikov-Thorne1973}) profiles on the equatorial plane for both a NkS and a BH. The Keplerian angular momentum has a minimum close to the BH, i.e., any angular momentum lower than that value can reach the event horizon. However, for a NkS, we do not see any such minimum for the given spin parameter ($a_*=1.01$), and it keeps on decreasing. This essentially suggests that accretion with given angular momentum can not reach the central NkS. We expect the accretion flows to bounce back due to the reflective nature of the potential.
\begin{figure}
    \centering
    \includegraphics[width=0.45\textwidth]{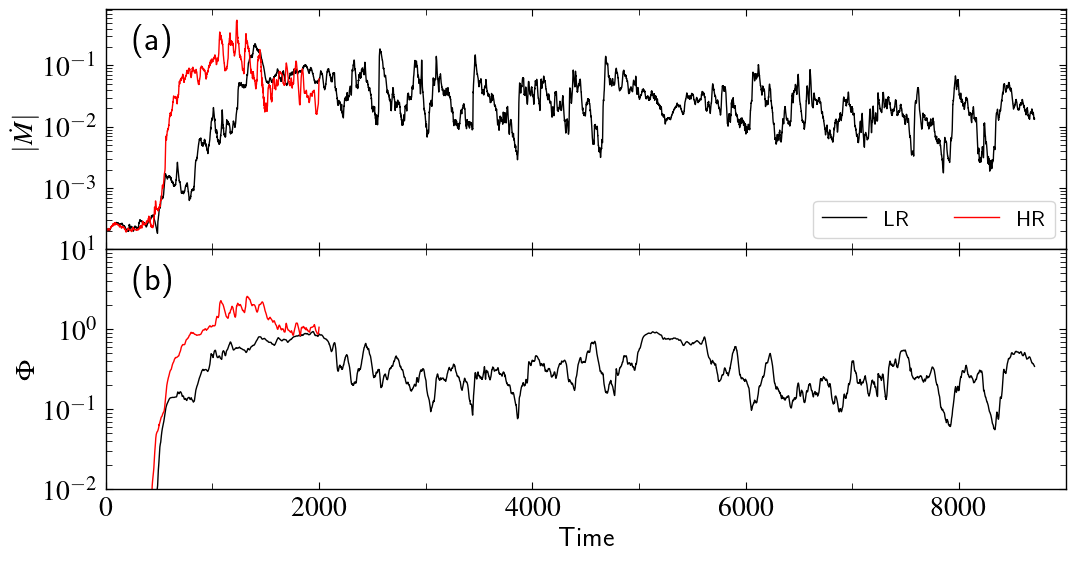}
    \caption{Same as Fig.~\ref{fig:Fig-02}(a,b), but comparison is shown for 3D low-resolution (LR) and high-resolution (HR) runs.}
    \label{fig:Fig-A2}
\end{figure}

\begin{figure}
    \centering
    \includegraphics[width=0.45\textwidth]{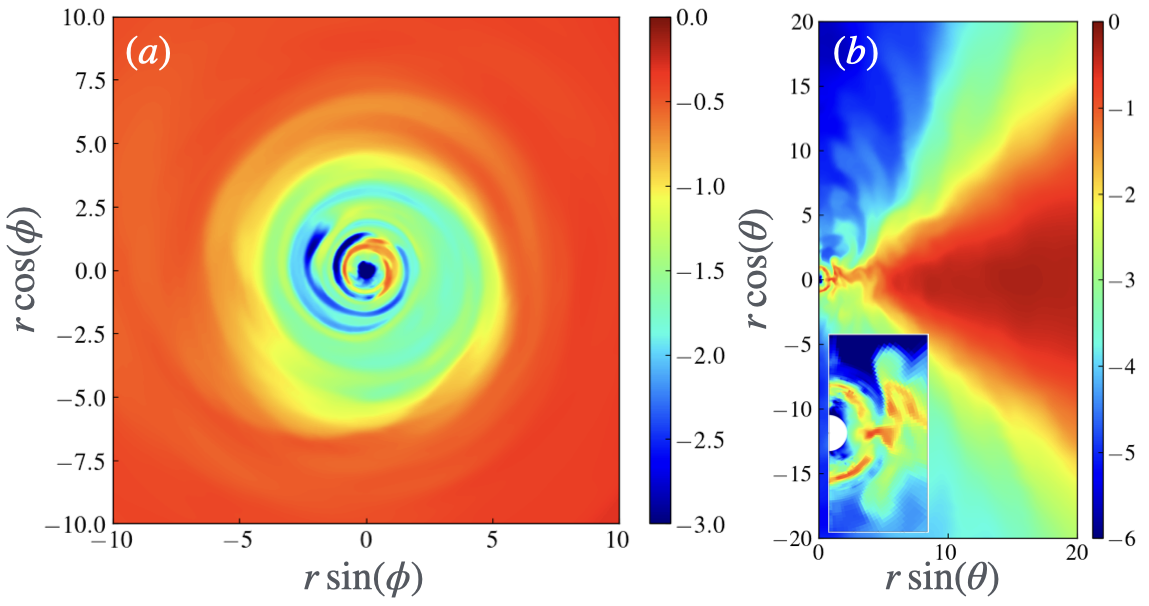}
    \caption{Same as Fig.~\ref{fig:Fig-06}(a,b), but for low-resolution model (M$1.0$R3DLR) at time $t=8000\,t_g$.}
    \label{fig:Fig-A3}
\end{figure}

\section{Comparison with long term low-resolution 3D results}\label{appendixD}
In this section, we compare the results from high-resolution (M$1.0$R3DHR) and low-resolution (M$1.0$R3DLR) 3D NkS cases. In Fig.~\ref{fig:Fig-A2}, we show the same variables as seen in Fig.~\ref{fig:Fig-02}(a,b). The panels suggest that the variability in the mass flux (inflow and jet/outflow) has longer timescales of variability than that of the axisymmetric runs. The values of $|\dot{M}|$ and $\Phi$ reach a quasi-steady state for simulation time $t>1500\,t_g$ for both resolutions. Finally, in Fig.~\ref{fig:Fig-A3}(a,b), we show the same as seen in Fig.~\ref{fig:Fig-06}(a,b), but for low-resolution model (M$1.0$R3DLR) at time $t=8000\,t_g$. These panels confirm that the high-resolution earlier time results hold consistent even for low-resolution long-term simulations. The distinguishable feature, i.e., a quasi-spherical density distribution around the NkS, can be seen in the later simulation time.
\bibliography{references}
\end{document}